\documentclass[journal=apchd5,manuscript=letter,layout=preprint]{achemso}

\usepackage{todonotes}
\usepackage[version=3]{mhchem} 

\usepackage{amssymb}
\newcommand*\real{\operatorname{Re}}

\author{Emerson G. Melo}
\affiliation[University of Campinas]{Applied Physics Department, Gleb Wataghin Physics Institute, University of Campinas, Campinas, SP, Brazil}
\alsoaffiliation[Photonicamp]{Photonics Research Center, University of Campinas, Campinas, SP, Brazil}
\altaffiliation{Current address: Materials Engineering Department, Lorena School of Engineering, University of S\~{a}o Paulo, 12602-810 Lorena, SP, Brazil}
\author{Ana L. A. Ribeiro}
\affiliation[University of Campinas]{Applied Physics Department, Gleb Wataghin Physics Institute, University of Campinas, Campinas, SP, Brazil}
\alsoaffiliation[Photonicamp]{Photonics Research Center, University of Campinas, Campinas, SP, Brazil}
\author{Rodrigo S. Benevides}
\affiliation[University of Campinas]{Applied Physics Department, Gleb Wataghin Physics Institute, University of Campinas, Campinas, SP, Brazil}
\alsoaffiliation[Photonicamp]{Photonics Research Center, University of Campinas, Campinas, SP, Brazil}
\author{Antonio A. G. V. Zuben}
\affiliation[University of Campinas]{Applied Physics Department, Gleb Wataghin Physics Institute, University of Campinas, Campinas, SP, Brazil}
\alsoaffiliation[Photonicamp]{Photonics Research Center, University of Campinas, Campinas, SP, Brazil}
\author{Marcos V. P. Santos}
\affiliation[University of Campinas]{Applied Physics Department, Gleb Wataghin Physics Institute, University of Campinas, Campinas, SP, Brazil}
\author{Alexandre A. Silva}
\affiliation[Samsung Research Brazil]{Samsung Research Brazil, Av. Cambacica 1200, Parque dos Resedas, 13097-160 Campinas, SP, Brazil}
\author{Gustavo S. Wiederhecker}
\affiliation[University of Campinas]{Applied Physics Department, Gleb Wataghin Physics Institute, University of Campinas, Campinas, SP, Brazil}
\alsoaffiliation[Photonicamp]{Photonics Research Center, University of Campinas, Campinas, SP, Brazil}
\author{Thiago P. M. Alegre}
\email{alegre@unicamp.br}
\affiliation[University of Campinas]{Applied Physics Department, Gleb Wataghin Physics Institute, University of Campinas, Campinas, SP, Brazil}
\alsoaffiliation[Photonicamp]{Photonics Research Center, University of Campinas, Campinas, SP, Brazil}

\title[Structural Colors]{Bright and Vivid Diffractive-Plasmonic Structural Colors}

\begin{document}

\begin{tocentry}
	\includegraphics[scale=1.0]{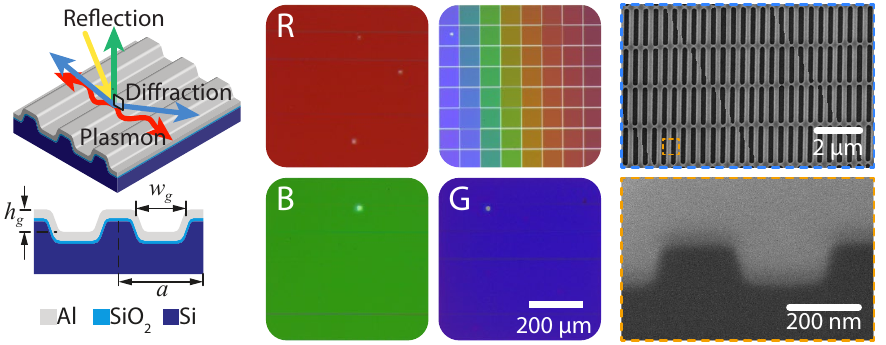}
\end{tocentry}


\begin{abstract}
	\textbf{Colors observed in nature are very important to form our perception of an object as well as its design. The desire to reproduce vivid colors such as those found in birds, fishes, flowers and insects has driven extensive research into nanostructured surfaces. Structural colors based on surface plasmon resonance (SPR) have played an important role in this field due to its high spatial resolution. Creating vivid color reflecting surfaces is still a major challenge and could revolutionize low power consumption image displays. Here we combine diffraction and plasmonic effects to design bright and vivid-color reflecting surfaces. The periodic reflection patterns are designed through genetic algorithm optimization and are defined in aluminum coated silicon chips.}
\end{abstract}

Development of pigments and dyes of specific colors has a long history as it improves the appearance of clothes, food, buildings, pictures, and so forth. These compounds absorb a range of light wavelengths and the reflected or transmitted bandwidth defines the color properties. This principle is the basis of most modern liquid-crystal color displays and printers~\cite{Jin.Woong2018,Jun.Choi2013,Horng-Show.Koo2006}. Despite the great technological achievements by the pigment- or dye-based display and imaging devices, their spatial resolution is still at micrometer scale, they can suffer from particle agglomeration, and  performance degradation when exposed to high temperature or ultraviolet radiation~\cite{Ram.W.Sabnis1999,Mehdi.K.Hedayati2017,Jin.W.Namgoong2018}; also, many chemical components present in colorant agents are toxic with limited reciclability~\cite{Mehdi.K.Hedayati2017}.

Advances in nanofabrication have opened new ways for novel structures, such as meta-surfaces and surface plasmon resonance (SPR) devices, that can enhance the control over light-matter interactions and lead to the creation of exciting new effects~\cite{Hou-Tong.Chen2016,Francesco.Monticone2017,Meinzer.Nina2014}. Through the interaction with nanopatterned structures, light can be scattered and absorbed in such a way to control its intensity, phase and polarization\cite{Ebbesen.T.W1998,Yu.Nanfang2011,Chen.Xianzhong2012}, mimicking the vibrant and vivid structural colors found in nature~\cite{Fu.Yulan2016, Kumar.Karthik2012,Ram.W.Sabnis1999,Mehdi.K.Hedayati2017}.
\begin{figure*}[ht]
	\centering
	\includegraphics[scale=1.0]{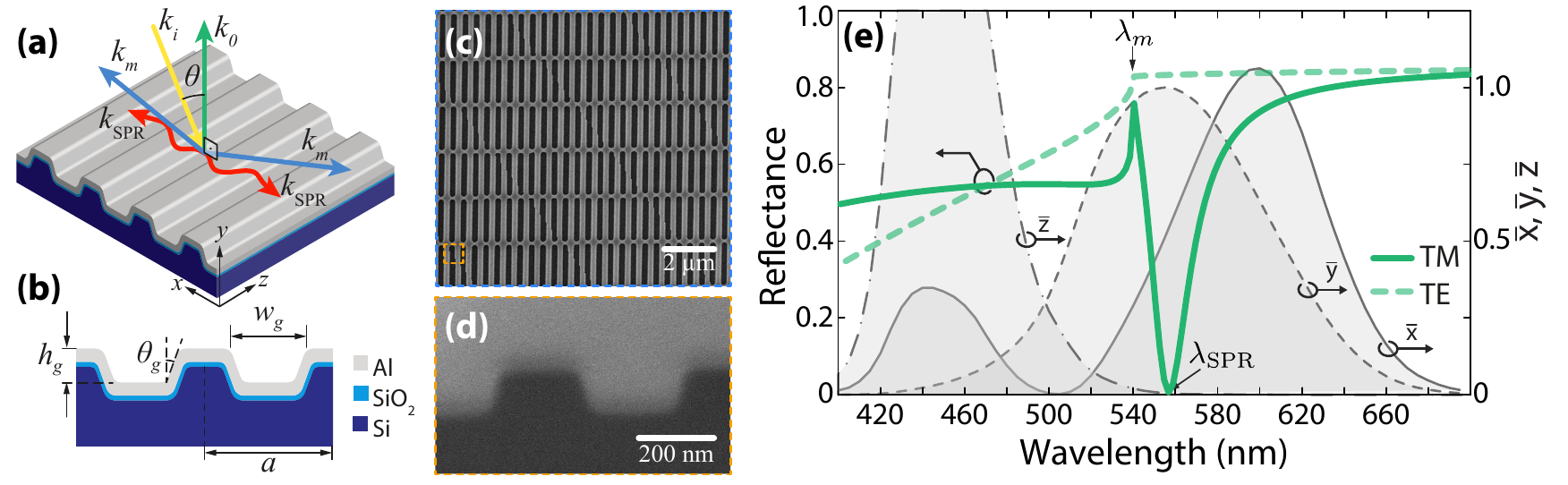}
	\caption{Grating Structure. (a-b) Schematic representation of the device structure composed of a nanopatterned Si substrate recovered by a metallic Al thin film. The wavelengths of the incident light $k_{i}$ can be diffracted $k_{m}$ ($m$ is an integer) or absorbed ($k_\text{SPR}$) depending on the grating geometry defined by lattice constant $a$, the groove height $h_{g}$, width $w_{g}$, and sidewall slope $\theta_{g}$. A native oxide layer is present above the Si surface; (c-d) Scanning electron microscopy (SEM) images showing the top and cross-sectional views of a fabricated device. (e) Calculated reflection spectra (green) for TE (dashed) and TM (solid) polarization of a shallow grating superposed to the CIE 1931 $\bar{x}$,$\bar{y}$,$\bar{z}$ standard observer color matching functions. The diffraction cutoff ($\lambda_m$) and surface plasmon resonance ($\lambda_\text{SPR}$) wavelengths are also seen.}
	\label{fig1:device}
\end{figure*}
Among the routes of structural color creation~\cite{Lee.Taejun2018}, SPR color purity and spatial resolution stands out\cite{Mehdi.K.Hedayati2017,Kristensen.Anders2016,Gu.Y2015}. The SPR can confine the optical excitation along a metal-dielectric interface far below the diffraction limit, allowing the exploration of structures with one or two orders of magnitude smaller than the traditional pigment-based or dye-based color filters \cite{Kumar.Karthik2012}. 

Plasmonic color filters working under transmission mode have been extensively studied \cite{Davis.Matthew2017,Mahani.F.Fouladi2017,Djalalian-assl.Amir2014,Ang.B.O.F2017,Li.Zhongyang2015,Xu.Ting2010,Lee.S.Uk2017,WILLIAMS.C.2016,Park.Chul-soon2015}, such devices are placed between the panel screen and a back-light source, thus blocking all the wavelengths outside their color bandwidths. 
However, when the device is intended to work under an ambient light illumination, narrowband reflection filters are desired, a far more challenging configuration. Some nanostructured devices based on diffractive coupling effects or enhanced resonances can exhibit such narrowband reflection, but only at large incident angles~\cite{Olson.Jana2014,Olson.Jana2016,Fan.J.R.2017,Guangyuan.Si2014}. Despite various attempts to overcome these angle limitations, the obtained colors are still far from the standard color systems~\cite{Tan.S.J2014,Lochbihler.Hans2018,Wang.Hao2017,Fatemeh.F.Mahani2018,Williams.Calum2017,Mahani.F.Fouladi2017,Kumar.Karthik2012,Wang.Liancheng2016,Mudachathi.Renilkumar2017,Goh.X.Ming2014}.

Here, we show how to combine two distinct physical effects to create narrow band reflection filters. Controlling both the first order diffraction and the surface plasmon resonance, we created nanopatterned surfaces which are able to reproduce bright and vivid structural colors along the whole visible spectrum. Furthermore, we employ a genetic algorithm to obtain the three primary colors (red, green, and blue) that form the basis of a large gamut color system. The devices presented here are suitable to fit both display and imaging requirements such as high resolution pixel sizes, high color purity, and good brightness performance.

\begin{figure*}[ht!]
	\centering
	\includegraphics[scale=1.0]{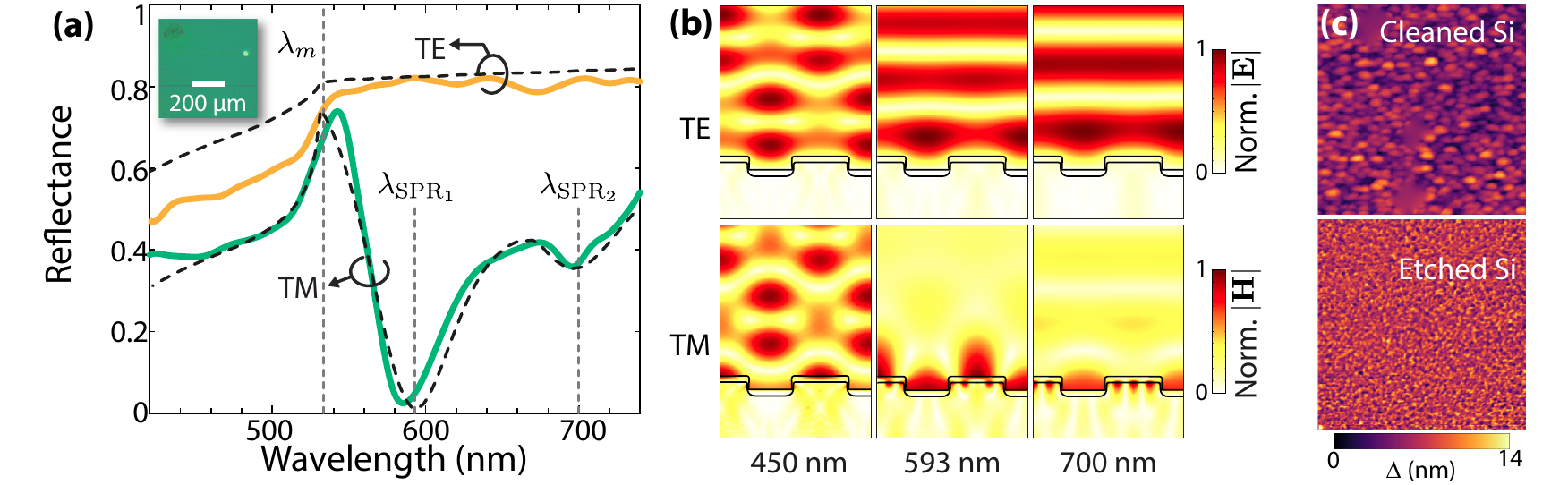}
	\caption{Metal Parameters Fitting. a) Measured (solid) and fitted (dashed) reflectance spectra for TE and TM polarization for a sample. The fitting was done using a multiobjective GA method for obtain the Drude-Lorentz parameters of the Al dielectric functions.  $\lambda_{m}$ ($\lambda_\mathrm{SPR_1}$ and  $\lambda_\mathrm{SPR_2}$) is the Rayleigh (Wood) anomaly wavelength. Inset: Optical microscopy image of the test grating structure with lattice $a=530$~nm, grove height $h_g=80$~nm, and near vertical sidewall slope. (b) Intensity distribution of the electric and magnetic fields calculated for TE and TM input polarizations at wavelengths bellow the diffraction cutoff, and at the two surface plasmon resonances ($\lambda_\mathrm{SPR_1}$ and  $\lambda_\mathrm{SPR_2}$), respectively. (c) Atomic Force Microscopy (AFM) images of $30$~nm Al thin film deposited over a clean and an etched Si wafer. The average RMS roughness of both sample is $1.9$~nm.
	}
	\label{fig2:fitting}
\end{figure*}

A schematic view of the device proposed in this work is illustrated in Fig.~\ref{fig1:device}(a). When the incident light ($k_i$) interacts with the grating, it can be scattered into different diffraction orders ($k_m$), but it also excites a SPR mode ($k_\text{SPR}$), being absorbed by the metal film. In Fig.~\ref{fig1:device}(b), the cross-sectional representation of the device shows a nanopatterned Si substrate covered by a thin native SiO$_{2}$ layer and an Al thin film. Standard nanofabrication processes and CMOS compatible materials were used to fabricate such structures. The grating lattice constant $a$ and groove width $w_{g}$ were defined by electron-beam lithography, whereas both the groove height $h_{g}$ and sidewall slope $\theta_{g}$ were controlled by a Si plasma-etching process. After that, a thin Al film ($\approx$32~nm thick) was deposited through electron-beam evaporation. Scanning electron microscopy (SEM) images of top and cross-sectional views of the fabricated structure are shown in Fig.~\ref{fig1:device}(c-d), respectively. Further details on the fabrication process can be found on Methods section. 

The optical response of an ordinary metallic grating, as the one seen in Fig.~\ref{fig1:device}(e), exhibits two distinct physical effects for transverse magnetic (TM) polarized light, namely, plasmonic resonance and diffraction, through which vivid colors could be created. For that, the reflectance spectrum must be molded in such a way which suitable color properties are obtained when multiplying it and the $\bar{x}\left(\lambda\right)$, $\bar{y}\left(\lambda\right)$, and $\bar{z}\left(\lambda\right)$ color matching functions. The plasmonic resonance position can be estimated by solving the wave equation for a metal-dielectric interface, together with energy and momentum conservation. The SPR excitation wavelength can be calculated as \cite{Maystre2012}:
\begin{equation}
\label{eq:wood}
\lambda_\text{SPR} = \frac{a}{m}\left(\pm\real\left\lbrace \sqrt{\frac{\varepsilon_d\varepsilon_m(\omega)}{\varepsilon_d + \varepsilon_m(\omega)}}\right\rbrace - \sqrt{\varepsilon_d}\sin\theta\right),
\end{equation}
where $\varepsilon_m(\omega)$ is the metal permittivity as a function of the incident light frequency $\omega$, $\varepsilon_d$ is the dielectric constant of the medium adjacent to the metal, $m$ is a integer related to the diffraction order, and $\theta$ is the light incident angle from the normal direction of the grating surface. Despite this model simplicity, when using the dielectric values from the metal and air, it predicts the experimentally measured plasmon resonance dip position $\lambda_\mathrm{SPR_1}$ on Fig.~\ref{fig2:fitting}(a) within 10\% of the observed value. Furthermore, it also provides a hint that a second longer wavelength plasmonic resonance related to the metal-oxide interface may occur around $\lambda_\mathrm{SPR_2}$, as really seen around 700 nm wavelength.

The other remarkable feature of the reflected light spectrum, which can be observed for both the transverse electric (TE) and the TM light polarization, occurs due to diffraction. Below a certain wavelength, normal reflection can be heavily suppressed, as shown in Fig.~\ref{fig1:device}(e), since light is coupled to diffraction modes which travel in shallow angles with respect to the grating surface. The diffraction wavelength cutoff can be predicted using the so-called Rayleigh's relation \cite{Maystre2012}:
\begin{equation}
\label{eq:rayleigh}
\lambda_{m} = \frac{a\sqrt{\varepsilon_d}}{m}(\pm 1 - \sin\theta),
\end{equation}
where $\lambda_{m}$ corresponds to the cutoff wavelength of the $m^\text{th}$ diffractive mode. This equation shows that the Rayleigh wavelength is dictated by the grating geometry and the incident medium, not relying on the metal characteristics. For near normal incident angle ($\theta\approx 0$) in air, Eq.~(\ref{eq:rayleigh}) resumes to $\lambda_{m} = a$ for first order diffraction modes ($m = \pm1$). Such approximation predicts the cutoff wavelength in Fig.~\ref{fig2:fitting}(a) within 5\% difference regarding the observed value.

\begin{figure*}[ht!]
	\centering
	\includegraphics[scale=1.0]{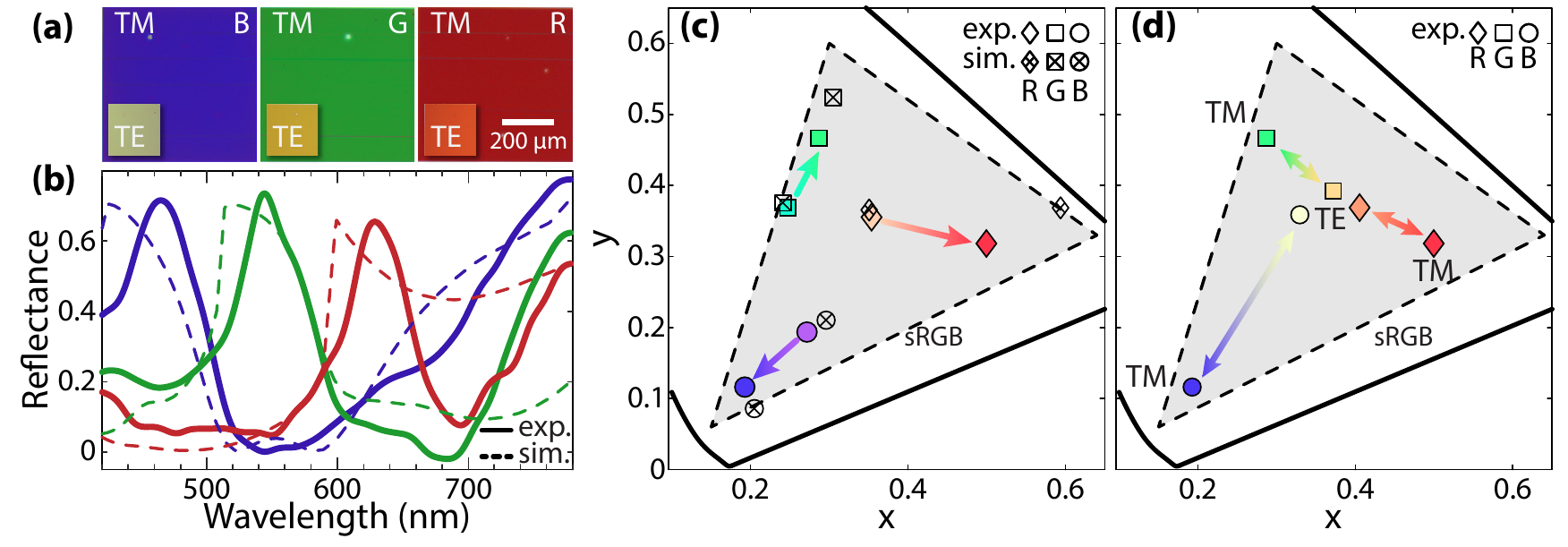}
	\caption{Optimization Strategy. (a) Measured microscope images for the blue, green, and red color pixels fabricated using the geometric parameters obtained by the GA optimization. (b) Measured (solid lines) and simulated (dashed lines) reflectance spectra for each of the pixels shown in (a). (c) CIE 1931 Chromaticity Diagram showing the color coordinates of the measured and simulated grating structures. The arrows indicate the changes of the
Changed more than to a
Sep 25, 2019 1:19 AM • You
6
 measured color coordinates after the GA optimization, which results in a four-fold improvement with respect the gamut area coverage of the test sample. (d) Changes in measured color coordinates for the optimized devices when the polarization is rotated from TM to TE.}
	\label{fig3:ga_result}
\end{figure*}

Although the metal dielectric function does not affect the diffractive cutoff wavelength, it is an essential information during the geometry optimization that accounts for both diffraction efficiency and SPR. The metal dielectric function is significantly impacted by the root-mean-square (RMS) roughness and the lateral correlation length~\cite{Timalsina.Y.P.2015}. The statistical analysis of the Atomic Force Microscopy (AFM) images (Fig.~\ref{fig2:fitting}(c)) reveals the same RMS roughness but different lateral correlation lengths for the Al thin films deposited over a pristine or etched Si surface. We use this information to extract the Drude-Lorentz parameters (Table S1) from a reflectance fitting model which took into account the measured optical response of three different color pixels regarding a non optimized test sample (see Fig. S1 and the discussion therein). The fitting yields a damping frequency for the etched surface (grooves) roughly  1.7 times larger than the pristine  surface (ridge), in agreement with previously  measured roughness correlation lengths~\cite{Timalsina.Y.P.2015, Yi-Cheng.C.2017}. The fitted data allows to estimate the grating response for both light polarizations using numerical simulations. Fig.~\ref{fig2:fitting}(b) shows the resulting simulated distribution of the electric and magnetic fields calculated for TE and TM input polarizations. For wavelengths below $\lambda_{m}$, light is coupled into the first diffraction mode, thus reducing the specular reflected power. Above $\lambda_{m}$, a drop in the reflectance spectrum is observed around $600$~nm only for TM polarization as a consequence of the excitation of a strong SPR resonance. The numerical simulation also shows that the longer wavelength plasmon around 700 nm is located mainly at the metal-substrate interface, as previously predicted. The oxide layer is responsible for reduces the effective index and brings $\lambda_\mathrm{SPR_2}$ into lower wavelengths, thus creating a larger absorption bandwidth (see Fig. S2 for more discussion on this second dip). This may actually improve the color purity of greenish or blueish pixels as it further suppress the reflection of red spectral portion.

A fine tune of the grating geometrical parameters can control its diffractive-plasmonic like optical response, so that a sharp reflectance peak is produced. Such narrow peak will be ultimately responsible for creating the vivid colors in our structures. The analytical models of Eqs. \ref{eq:wood} and \ref{eq:rayleigh} provide a guideline for engineering the narrow reflection peak position and width, however, a realistic design must take into account the exact geometric structure and the full complexity of Maxwell's equation. We perform a Genetic Algorithm (GA) optimization to accounts for all the grating geometry degrees of freedom, in order to obtain suitable geometric parameters, thus enabling the brightest and purest primary RGB colors. Employing the GA optimizations was critical for the final results, since achieving saturated colors for each primary RGB colors requires individual tuning of the diffraction and SPR resonances. For instance, obtaining a saturated blue (B) pixel requires critical coupling condition (high extinction) of the SPR for wavelengths above $500$~nm. Meanwhile, saturated red (R) pixels do not depend so critically on the SPR resonance extinction, because the wavelength response of our eyes will naturally provide a high-pass filter for the near-infrared light. In this case, increasing the coupling of light into the first diffraction modes over wavelengths below $620$~nm is enough to ensure  vivid red colors. Finally, obtaining a green (G) color is only possible by finding a fine balance between light coupling into diffraction and SPR modes. 

\begin{figure*}[ht]
	\centering
	\includegraphics[scale=1.0]{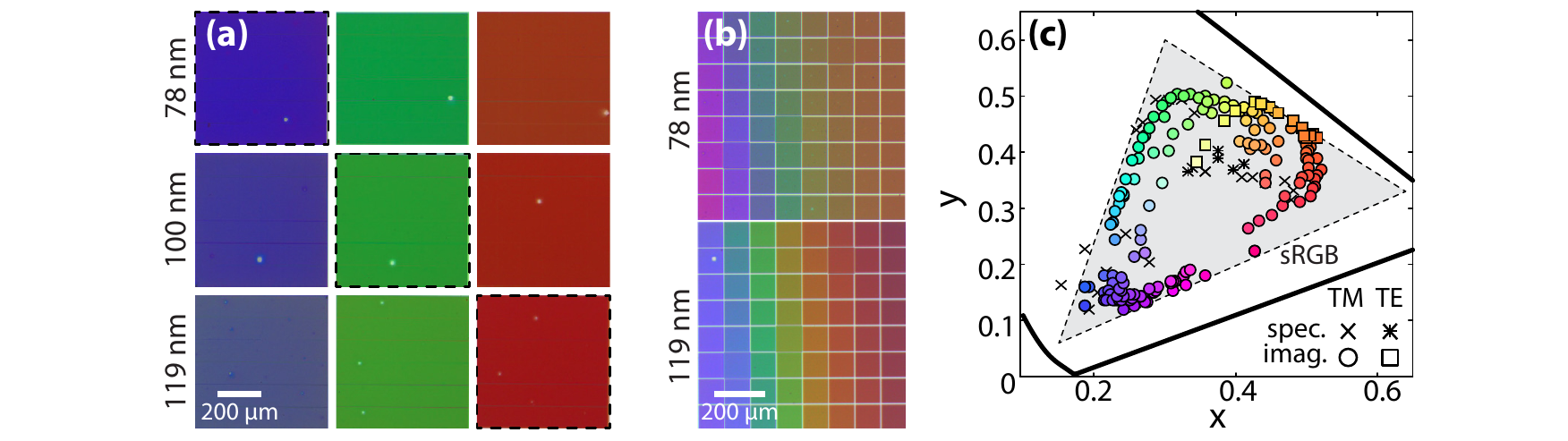}
	\caption{Achieved Color Gamut. (a) Optical microscopy images of individual pixels with different $h_g$ values. (b) Optical microscopy images of color palettes fabricated by adjusting the lattice constants of the optimized grating structures from $400$~nm up to $700$~nm. (c) CIE 1931 Chromaticity Diagram including the $xy$-coordinates calculated from the measured reflectance spectra as well the ones obtained from color palettes images.}
	\label{fig4:gamut}
\end{figure*}

Fig.~\ref{fig3:ga_result}(a) shows a microscope picture of each of the optimized structure for both TM and TE incident polarized lights. Fig.~\ref{fig3:ga_result}(b) presents the measured (solid) and simulated (dashed) reflectance spectra. The GA optimization has tailored the grating structures following the intuition previously discussed; the B pixel critically couples to a large bandwidth SPR; the R response exhibits the higher diffraction efficiency; whereas the G spectrum denotes a good compromise between these two characteristics. As a consequence, all the three curves present narrow linewidths ($\approx50$~nm) with high reflectivity ($>70$\%), which enhances both the brightness and color saturation. The full procedure as well as the resulting parameters from the GA optimization are discussed in Supplementary Material Sec.~S2.

In order to compare the results with and without the optimization, the xy color coordinates for each pixel before and after the optimization procedure (arrows) are shown in the CIE 1931 Chromaticity Diagram on Fig.~\ref{fig3:ga_result}(c). After the optimization process, a four-fold improvement is achieved with respect to the color gamut area coverage of the test sample.

The small differences between the $xy$ coordinates of the experimentally measured colors with respect the simulated ones can be attributed to geometric deviations introduced during the fabrication processes. For example, when compared with the simulated spectra, higher reflectance is observed at lower wavelengths in the G and R experimental spectra, whereas lower reflection is observed above the Rayleigh's wavelength. Irregularities on grating surface can result in such situation by scattering the diffracted light back into the fiber probe, and improving the SPR coupling~\cite{Endriz.J.G.1971}. The larger mismatch between the measured and simulated R color coordinates is attributed to its higher dependence on diffraction efficiency. A sensitivity analysis was performed and reveals that deviations of grating height ($h_{g}$) are responsible for the larger changes in color properties, as can be seen in Fig.~\ref{fig4:gamut}(a). Nevertheless, the structures show a very good resilience with respect the groove sidewall slope (Fig. S6).

The optical response of gratings with patterns made of straight lines is very sensitive to polarization. This feature can be explored for saturation color tuning by changing the polarization of the incident light between TM and TE states, as represented in Fig. \ref{fig3:ga_result}(d) by the colored arrows. Images of the samples taken under TE light exposure are shown as insets on the TM images of Fig. \ref{fig3:ga_result}(a).  The only color that is rather insensitive to polarization is the red one, as it  weakly  relies on the plasmonic resonance.

An example of the colors that can be generated by varying only the lattice constants of the optimized grating structures are shown on Fig.~\ref{fig4:gamut}(b). The color palettes can be gradually tuned from a vivid blue to red colors --  passing through green --  when the lattice parameter varies from $400$~nm to $700$~nm. Fig. \ref{fig4:gamut}(c) shows the CIE 1931 Chromaticity Diagram including the $xy$ color coordinates calculated from the measured reflectance spectra, as well as the ones obtained from the camera-captured images. The resultant color gamut covers a large region of the sRGB gamut delimited by the gray area.  


In conclusion, using standard nanofabrication processes and CMOS compatible materials, we have fabricated reflective color filters based on Al thin film over a nanopatterned Si substrate. The optical response of these devices, characterized both through reflectance spectra and optical microscopy images, show that it is possible to generate bright and vivid structural colors relaying on a suitable control of the first order diffraction modes as well as the surface plasmon dispersion. The usage of a Genetic Algorithm optimization process proves to be efficient to tailor the grating geometry in order to achieve specific colors along the whole visible spectrum, including, but not limited to, the three primary colors: red, green, and blue. The optimized structures result in an four-fold improvement with respect to the color gamut area coverage of non optimized devices. The devices presented here can be employed as color filter pixels for reflective colored displays with low power consumption and high image resolution, in which the viewing angle is not a concern such as handheld devices, as well as for color printing of static images in micrometer scale for security applications, angle dependent colored images for decoration purposes, or color sensor and optical filters.

\section{Methods}

\subsection{Fabrication}
The reflective color pixels were fabricated by employing standard nanofabrication techniques and CMOS compatible materials, such as Si and Al. Firstly, the positive electron-beam resist ZEP520A (ZEON$^{TM}$ Coorp) was spin-coated for 1 minute onto a cleaned Si (100) wafer at 6000 RPM. The grating patterns were then exposed in the electron-beam resist thin film in a Raith E-Line Plus$^{\textcopyright}$ system using 70 $\mu$C/cm$^{2}$ area dose and 30kV acceleration voltage. The grating profiles were then transferred to the Si substrate by Inductively-Coupled Plasma Reactive-Ion Etching (ICP-RIE) using a gas mixture containing C$_{4}$F$_{8}$, SF$_{6}$, and Ar on an Oxford Plasmalab$^{\textcopyright}$ 100 ICP Etcher. Afterwards, an O$_{2}$ plasma was employed in order to remove small electron-beam resist residues which may be bonded on Si surface after the usual cleaning process. Thus, a $\approx$32 nm thick Al thin film was deposited onto the device surface using electron-beam evaporation.

\subsection{Optical Characterization}
The reflectance spectra from the fabricated samples were measured using a Thorlabs$^{TM}$ spectrometer which receives light collected from a multi-mode fiber bundle. The samples were directly illuminated with a xenon-mercury lamp using a fiber probe adjusted for normal light incidence angle. The color filter power spectrum was averaged over 50 measurements with an integration time of 100 ms. The spectra of flat Al surfaces nearby the grating structures were employed for normalization purposes. A polarization control placed between the sample and the fiber probe allowed to excite the color filters on both TM and TE modes. The final reflectance spectra were obtained after the usage of a 70th-order one-dimensional median filter followed by a 3rd-order low pass filter, which were employed to filter out some jitters and peaks, caused by the mercury lines, from the raw data (Fig S7). Micropositioner stages with angular control were used to align the fiber above the samples with near normal direction with respect the grating surface. The sample images were obtained by a Thorlabs$^{\textcopyright}$ CCD camera using a 5x objective lens.

\subsection{Numerical Simulations}
Numerical simulations were performed with the Finite Elements Method (FEM) --- Comsol Multiphysics$^{\textcopyright}$ 5.3a --- using a two-dimensional grating unit cell and periodic (Bloch-Floquet) boundary conditions. The mesh size was extremely refined at the grating region in order to obtain a smooth transition of the fields across the media interfaces, thus ensuring a steady optical response. The Genetic Algorithm optimization process and data post-processing were made using Matlab$^{\textcopyright}$ software package.


\begin{acknowledgement}

This work was supported by Minist\'erio da Ci\^encia, Tecnologia, Inova\c{c}\~oes e Comunica\c{c}\~oes (MCTI), Coordena\c{c}\~ao de Aperfei\c{c}oamento de Pessoal de N\'ivel Superior - Brasil (CAPES) - Finance Code 001, National Counsel of Technological and Scientific Development (CNPq), and S\~ao Paulo Research Foundation (FAPESP) through grants: 2018/15580-6, 2018/15577-5, 2016/18308-0, 2012/17610-3 and 2012/17765-7. Part of results presented in this work were obtained through the 4716-11 project, funded by Samsung Eletrônica da Amazônia Ltda., under the Brazilian Informatics Law 8.248/91 The authors thank the Center for Semiconductor Components and Nanotechnologies (CCSNano) for the nanofabrication infrastructure.

\end{acknowledgement}

\begin{suppinfo}

Data and simulation files are available online\cite{Melo.E.G.2019} at Zenodo repository.

\end{suppinfo}


\footnotesize
\bibliography{diff-plasmon.bib}
\end{document}